\documentclass[12pt]{article}
\usepackage{amsmath}
\usepackage{amsfonts}
\usepackage{amssymb}
\usepackage{mathrsfs}
\usepackage{multicol}
\usepackage{color}
\usepackage{graphicx}
\usepackage{pstricks}
\usepackage{pst-node}
\usepackage{wrapfig}
\usepackage{enumerate}
\usepackage{caption}
\usepackage{cite}
\usepackage{subfigure}
\usepackage{amsfonts,amssymb}
\usepackage{amsthm}
\usepackage{amscd}
\definecolor{gray1}{gray}{0.1}
\definecolor{gray2}{gray}{0.2}
\definecolor{gray3}{gray}{0.3}
\definecolor{gray4}{gray}{0.4}
\definecolor{gray5}{gray}{0.5}
\definecolor{gray6}{gray}{0.6}
\definecolor{gray7}{gray}{0.7}
\definecolor{gray8}{gray}{0.8}
\definecolor{gray9}{gray}{0.9}

\newpsobject{showgrid}{psgrid}{subgriddiv=1,griddots=10,gridlabels=6pt}

\definecolor{dark-green}{rgb}{0,0.7,0}
\definecolor{dark-blue}{rgb}{0,0.2,0.5}
\definecolor{med-blue}{rgb}{0,0.7,1}
\definecolor{mblue}{rgb}{0,0.2,1}
\definecolor{cnc}{rgb}{0.8,0,0}
\definecolor{light-red}{rgb}{1,0.8,0.8}
\definecolor{dark-yelow}{rgb}{1,0.8,0}
\definecolor{light-blue}{rgb}{0.8,0.9,1}
\definecolor{verylight-blue}{rgb}{0.93,0.95,1}
\definecolor{light-yelow}{rgb}{1,0.9,0.8}
\definecolor{grey}{gray}{0.88}

\newpsobject{showgrid}{psgrid}{subgriddiv=1,griddots=10,gridlabels=6pt}

\begin{document}

\thispagestyle{empty}

\setlength{\abovecaptionskip}{10pt}

\begin{center}
{\Large\bfseries\sffamily{Real scalar field kink(-antikink)s and their perturbation spectra \\
in a closed universe}}
\end{center}
\vskip 3mm
\begin{center}
{\bfseries{\sffamily{Betti Hartmann$^{\rm 1, 2}$, Gabriel Luchini$^{\rm 3}$, Clisthenis P. Constantinidis$^{\rm 3}$, Carlos F.S. Pereira$^{\rm 3}$}}}\\
\vskip 3mm{
{$^{\rm 1}$\normalsize{Instituto de F\'isica de S\~ao Carlos (IFSC), Universidade de S\~ao Paulo (USP), \\ CP 369,
13560-970 , S\~ao Carlos, SP, Brazil}\\
$^{\rm 2}$\normalsize{Department  of  Theoretical  Physics,  University  of  the  Basque  Country  UPV/EHU,  48080  Bilbao,  Spain}\\
$^{\rm 3}$\normalsize Departamento de F\'isica,
Universidade Federal do Esp\'irito Santo (UFES),\\
CEP 29075-900, Vit\'oria-ES, Brazil}}
\end{center}

\begin{abstract} 
In this paper we demonstrate that solitons of a simple real scalar field model that are {\it static and linearly stable} do exist when considered in
a (3+1)-dimensional, spatially compact space-time background, the static Einstein universe, which is a good approximation to the observed universe for sufficiently small time intervals. We study the properties of  these solutions  for a $\Phi^4$-potential and demonstrate that next to the fundamental solutions, excited configurations exist.  We also investigate general perturbations about the solitons, determine their eigenfrequency spectra and compare them with those of the perturbations about the vacua of the model. We find that the degeneracy with respect to the multipoles of the perturbation, which is present for the vacua, does no longer exist in the presence of the soliton. Moreover, specific perturbations correspond to zero modes of the system. 
Our results have applications
in condensed matter physics as well as computations of quantum effects (e.g. the Casimir energy) in spatially compact space-times in the presence of soliton-like objects.
\end{abstract}

\section{Introduction}
After the discovery \cite{Chatrchyan:2012xdj,Aad:2012tfa} of the Englert-Brout-Higgs-Guralnik-Hagen-Kibble (EBHGHH) boson \cite{brout,higgs,guralnik} in 2012, a fundamental scalar field in nature, the study of scalar fields in diverse
contexts of physics has become of interest again. Scalar fields had previously been present in a number of high and low energy physics
models such as in the theory of inflation in the primordial universe that is supposed to be driven by a slow-roll scalar field, as well
as in condensed matter, where collective phenomena are often effectively described by a scalar field. The typical example for the latter
case is the Ginzburg-Landau model \cite{Ginzburg:1950sr} describing superconductivity.

So-called ``solitonic solutions' of these field theory models are of particular interest, especially due to their possible interpretation as relativistic particles\cite{Manton_2008}. The existence of such configurations is limited by some important features of the model, e.g. its integrability\cite{PhysRevD.14.1524} and/or the existence of a topological degree characterizing the field as a topological soliton\cite{manton2004topological}. Besides, important restrictions are imposed by the dimensionality of space and the types of fields which are present in the model with the possibility of finding a static solitonic solution; this is given by Derrick's theorem\cite{Derrick:1964ww}. For theories in $(d+1)$-dimensional Minkowski space-time with only scalar fields and no terms involving derivatives higher than second order, this theorem predicts that only in $d=1$ or $d=2$ static solitons can be found. However, even in flat space-time, there are ways to evade this theorem, as in the case of Q-balls\cite{Lee:1991ax,Coleman:1985ki,Bowcock_2009} which are non-topological solitonic scalar field configurations that exist in more than one spatial dimension and possess a time-dependent phase. In $(3+1)$-dimensional Minkowski space-time, the scalar field model needs (at least) a 6-th order self-interaction potential for Q-balls to exist\cite{Coleman:1985ki, Volkov:2002aj, Kleihaus:2007vk,Kleihaus:2005me}. When extending these models to curved space-time, self-gravitating counterparts of $Q$-balls, so-called {\it boson stars} exist \cite{Kaup:1968zz,Ruffini:1969qy,Schunck:1999pm,Friedberg:1986tp,Jetzer:1991jr, Kleihaus:2007vk, Kleihaus:2005me}. In curved space-time, no general direct extension of Derrick's theorem seems to restrict the solitonic solutions and the possibility of finding them is open to be explored, although there has been recent progress for asymptotically
flat space-times \cite{Carloni:2019cyo}. 

In this paper, we study a real scalar field model in a $(3+1)$-dimensional Einstein universe which is spatially compact. This model has previously been studied in \cite{sudarsky}, but only for one very specific case.
The Einstein universe describes a  universe filled with a perfect fluid that has positive energy density and vanishing pressure. In addition, there is a positive cosmological constant that allows the existence
of a static solution to the Einstein equation. While it is well known that the present universe
is accelerately expanding and as such contains a substantial amount of dark energy that can
mathematically be described by a positive cosmological constant, the Einstein universe is
often used as an approximation to the ``real universe'' for sufficiently small time intervals, which makes
e.g. computations of quantum states feasible (see e.g. \cite{ford2,ford1,Ford:1978zt,Herdeiro:2005zj}).
Moreover, the compactness of this space-time is of interest to condensed matter physics,
in particular to Bose-Einstein condensation (see e.g. \cite{Altaie:2001vv,altaie:2000,Parker:1991jg,Altaie:1978dx}) since it is well known that the geometry of a finite system plays an important role
in the thermal behaviour of the bosons and the actual condensation process. 

 Here, we extend the analysis of \cite{sudarsky} considerably to show that additional solutions with interesting properties exist. Moreover, we study general linear perturbations about these solutions
 and compute the eigenfrequency spectra. The outline of our manuscript is as follows~: we give the
model in Section \ref{sec:model}, while we discuss the numerical solutions of the model in Section \ref{sec:solutions}. In Section \ref{sec:stability}, we present our results on the eigenfrequency spectra 
of the perturbations about the solutions obtained in Section \ref{sec:solutions}. We conclude in Section \ref{sec:conclusions}.

\section{The model}
\label{sec:model}
We study a self-interacting scalar field model in (3+1) dimensions with action given by~:
\begin{equation}
\label{eq:action}
S=\int {\rm d}^4 x \sqrt{-g}{\cal L}=\int {\rm d}^4 x \sqrt{-g} \left( - \frac{1}{2}\partial_{\mu} \Phi^* \partial^{\mu} \Phi - V(\vert\Phi\vert)\right) \ \ \ \ , \ \ \ \ 
V(\vert\Phi\vert)=\frac{\lambda}{4}\left(\vert\Phi\vert^2 -\eta^2\right)^2  \ \ \ , 
\end{equation}
where $g$ denotes the determinant of the metric tensor $g_{\mu\nu}$ of the (3+1)-dimensional space-time background that we assume to be non-dynamical. The self-coupling constant $\lambda$ is chosen positive, such that the potential is positive-definite, i.e. has its lowest value $V\equiv 0$ at what we will refer to as ``the vacua'' in the following, i.e. at $\Phi\equiv \pm \eta$.  A priori, we allow the scalar field
to be complex valued and the star denotes hence complex conjugation.

The equation of motion following from the variation of the action (\ref{eq:action}) with respect to the scalar field reads~:
\begin{equation}
\label{eq:eom}
\frac{1}{\sqrt{-g}}\partial_{\mu}\left(\sqrt{-g}\partial^{\mu}\Phi\right) - 2 \frac{\partial V}{\partial \vert\Phi\vert^2} \Phi  = 0 \ \ \ \ \Rightarrow \ \ \ \  \frac{1}{\sqrt{-g}}\partial_{\mu}\left(\sqrt{-g}\partial^{\mu}\Phi\right) - \lambda \left(\vert\Phi\vert^2 - \eta^2\right)\Phi = 0 \ , 
\end{equation}
while we assume the space-time background to be fixed. In the following, we will choose 
the static Einstein universe \cite{einstein1} with metric~:
\begin{equation}
\label{eq:einstein_universe}
ds^2 = -dt^2 + R_0^2\left[d\chi^2 +\sin^2\chi\left(d\theta^2 + \sin^2\theta d\varphi^2\right)\right] \ \ \ \ , 
\end{equation}
where  the coordinates $(t,\chi,\theta, \varphi)$ have the following ranges~:
$t \in ]-\infty:\infty[$, $\chi\in [0:\pi]$, $\theta\in [0:\pi]$, $\varphi\in [0:2\pi[$. 
Constant $\eta$-sections are 3-spheres with unit radius, i.e. the topology of this compact space-time is 
$\mathbb{R}\times S^3$. The metric given by (\ref{eq:einstein_universe}) is a  Friedman-Lem\^aitre-Robertson-Walker (FLRW) space-time and a solution to the Einstein equation $G_{\mu\nu}+\Lambda g_{\mu\mu} = 8\pi {\rm diag}(\rho,-p,-p,-p)$, where $\rho=\Lambda/4$, $p=0$ are the energy-momentum components of  a perfect fluid with energy density $\rho$ and pressure $p$. Moreover, the cosmological constant $\Lambda=R_0^{-2}$ is positive.
While the observed universe is known to be accelerately expanding, the Einstein universe is normally considered a good approximation to the FLRW space-time in sufficiently small time intervals and has frequently been
considered in computations of the quantum vacuum energy in the universe (see e.g. \cite{ford1}).

In the following, we will demonstrate that next to the vacuum solution $\vert\Phi\vert_0\equiv \pm \eta$,
this model contains non-trivial, localized and static solutions that resemble kinks and anti-kinks.
In the following, we will hence assume that $\Phi(\vec{r},t) \sim \phi(\chi)$, i.e. that the scalar field
depends only on $\chi$. Note that
in Minkowski space-time the existence of real, static scalar field solutions of this simple model is forbidden by Derrick's theorem \cite{Derrick:1964ww}.
Moreover, in a curved space-time background, solutions don't exist if the space-time is asymptotically flat \cite{Carloni:2019cyo}. However, in the spatially compact space-time that we are studying here, the additional length scale -- the radius of the 3-sphere $R_0$ -- leads to the existence of
solitonic solutions.
This has been shown in \cite{sudarsky} for the $\Phi^4$-potential, albeit only for one specific value of the radius (in relation to the mass of the scalar field). Note that scalar field solitons on an $S^3$, i.e. in a compact space, have been studied previously also in a non-linear
sigma model \cite{Brihaye:1998zy}. 

With these assumptions, the non-vanishing components of the energy-momentum tensor 
$T_{\mu\nu} =g_{\mu\nu} {\cal L} - 2 \frac{\partial {\cal L}}{\partial g^{\mu\nu}}$ read
\begin{equation}
\label{eq:em_scalar}
T_t^t =T_{\theta}^{\theta}=T_{\varphi}^{\varphi}=-\frac{1}{2} \frac{\phi'^2}{R_0^2} - V(\phi)  \  \ , \ \ 
T_{\chi}^{\chi}=\frac{1}{2}\frac{\phi'^2}{R_0^2} - V(\phi) \ \ ,
\end{equation}
where the prime now and in the following denotes the derivative with respect to $\chi$. The energy density of the solutions is given by
$\varepsilon=-T^t_t$ and associating
the Noether charge related to the time-translation invariance to the total conserved energy $E$ of the solution, we can write 
\begin{equation}
E=\int {\rm d}^3 x \sqrt{-g} \varepsilon = 4\pi R_0^3 \int\limits_0^{\pi} {\rm d} \chi \sin^2\chi 
\left(\frac{1}{2} \frac{\phi'^2}{R_0^2} + \frac{\lambda}{4}\left(\phi^2 -\eta^2\right)^2\right)    \ .
\end{equation}
As is obvious from the expression, we do not need to require the solutions to tend to the
vacua of the potential in order to obtain finite energy solutions, which would be necessary in
Minkowski or asymptotically Minkowski space-time. In fact, we will show below that the value of $\phi(\chi=0)$ and $\phi(\chi=\pi)$, respectively, depends solely on the ratio between $R_0$ and the relevant length/mass scale in the model.

\section{Non-trivial solutions}
\label{sec:solutions}
Rescaling $\phi(\chi)\rightarrow \eta\phi(\chi)$, the equation of motion for the scalar field  reads 
\begin{equation}
\label{eq1}
\phi''+ 2\cot\chi \phi'- \alpha\phi(\phi^2 -1) = 0  \ ,
\end{equation}
where $\alpha=\lambda \eta^2 R_0^2$ is the only dimensionless parameter and corresponds
to the ratio between the radius of the 3-sphere and the core radius of the solution $\sim \left(\sqrt{\lambda} \eta\right)^{-1}$. 
The equation (\ref{eq1}) cannot be solved analytically requiring non-triviality. We have hence employed a numerical
technique \cite{colsys} and used the following boundary conditions in order to insure regularity of the solutions at $\chi=0$ and $\chi=\pi$~:
\begin{equation}
\label{eq:bc}
\phi'(\chi=0)=\phi'(\chi=\pi)=0 \ .
\end{equation}
The equation (\ref{eq1}) has been solved numerically in \cite{sudarsky}, however, only for a very specific value of $\alpha\equiv 4$.
In the following, we will discuss the behaviour of the solutions for generic values of $\alpha$ and refer to solutions 
tending from a positive value at $\chi=0$ to a negative value at $\chi=\pi$ as {\it antikinks}, while the corresponding
{\it kinks} are obtained easily by letting $\phi \rightarrow -\phi$, which is clearly a symmetry of (\ref{eq1}). 

The dimensionless energy density $\tilde{\varepsilon}$ and dimensionless energy $\tilde{E}$ of the solutions then reads
\begin{equation}
\tilde{\varepsilon}=\frac{1}{2}\phi'^2 + \frac{\alpha}{4}\left(\phi^2 -1\right)^2 \ \ \ , \ \ \ 
\tilde{E}=4\pi \int\limits_0^{\pi} {\rm d} \chi \sin^2\chi \ \tilde{\varepsilon}
\end{equation}
with $\varepsilon=\eta^2\tilde{\varepsilon}/R_0^2$ and $E=R_0 \eta^2 \tilde{E}$, respectively. \\
\\
Note that (\ref{eq1}) can be re-written in the following form~:
\begin{equation}
\label{eq:mechanical_analogue}
\frac{\phi'^2}{2} -\frac{\alpha}{4}\left(\phi^2 -1\right)^2 = C - 2\int\limits_0^{\chi} \cot\tilde{\chi} \phi'^2 \ {\rm d} \tilde{\chi} \ ,
\end{equation}
where $C$ is an integration constant. One can construct a mechanical analogue of this equation by considering $\chi$ to be ``time'' and $\phi$ the position describing a particle moving along a trajectory $\phi(\chi)$. Then the left-hand side of (\ref{eq:mechanical_analogue}) can be interpreted as the sum of kinetic and potential energy $U(\phi)=-\frac{\alpha}{4}\left(\phi^2 -1\right)^2$ of this particle with $C$ the ``energy'' and the integral term on the right-hand side a friction term.
\begin{figure}[ht!]
\begin{center}
\input{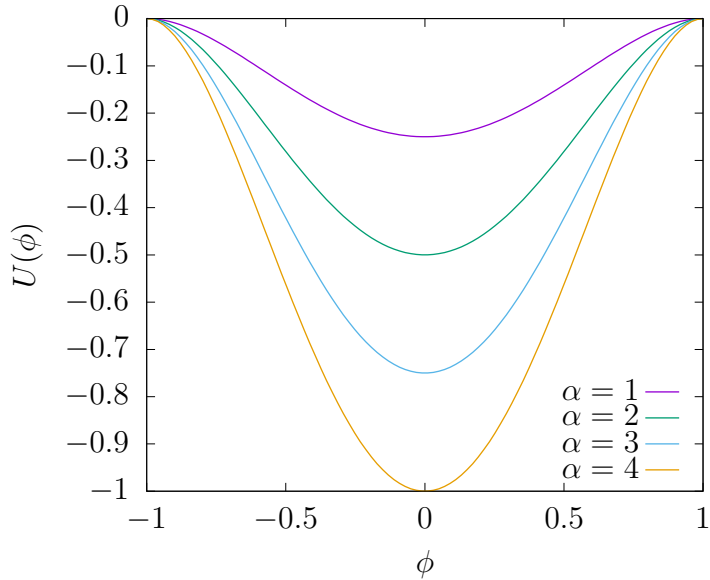}
\caption{We show the potential $U(\phi)=-\frac{\alpha}{4}\left(\phi^2 -1\right)^2$ (see discussion below (\ref{eq:mechanical_analogue})) in which the 
``particle'' would move on a ``trajectory'' $\phi(\chi)$ under the influence of a friction term for different values of $\alpha$.  }
\label{fig:potential_particle}
\end{center}
\end{figure}
The presence of this friction term is essential for the existence of the non-trivial solutions that we will present below. 
In Fig. \ref{fig:potential_particle}, we show $U(\phi)$ for different values of $\alpha$.  This indicates - taking the periodicity of the $
\chi$-direction into account - what type of solutions are possible. In contrast to flat space-time, the value of
$\phi(\chi)$ has to be ``fine-tuned'' with respect to $\alpha$ because $\alpha$ determines the slope of the potential, i.e. the value of $\phi'$ and hence the value of the friction term in (\ref{eq:mechanical_analogue}). Moreover, $\phi(\chi=0)\neq 1$ for $\alpha$ finite 
because we have the additional condition that $\phi(\chi=0)=\pm \phi(\chi=\pi)$ due to periodicity (see more details below).
In other words~: for the ``particle'' to roll back to $\pm \phi(\chi=0)$ in a given $\alpha$-potential (which possesses increased
slope when increasing $\alpha$) 
we have to choose $\phi(\chi=0)$ appropriately.
Moreover, Fig. \ref{fig:potential_particle} indicates that given enough ``initial energy'', i.e. starting at $\phi(\chi=0)$ sufficiently large,
the particle can role back to $\phi(\chi=\pi)=\phi(\chi=0)$ and, additionally, oscillate a number of times around $\phi=0$, i.e. we would
expect configurations whose shape resembles that of a pair of antikink-kinks as well as excited kinks and antikink-kinks, respectively, to exist as well. Our numerical results below show that these solutions do, indeed, exist in this model. It is clear that the existence of these solutions is fundamentally linked to the periodicity of the $\chi$-direction as well as the fact that we are in three spatial dimensions.

\subsection{Fundamental kinks and antikinks}
\label{subsection:kinks}

\begin{figure}[ht!]
\begin{center}
\input{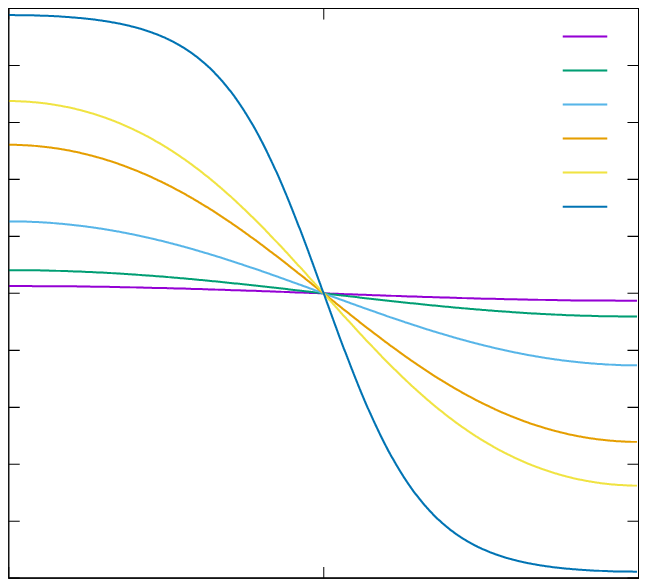}
\input{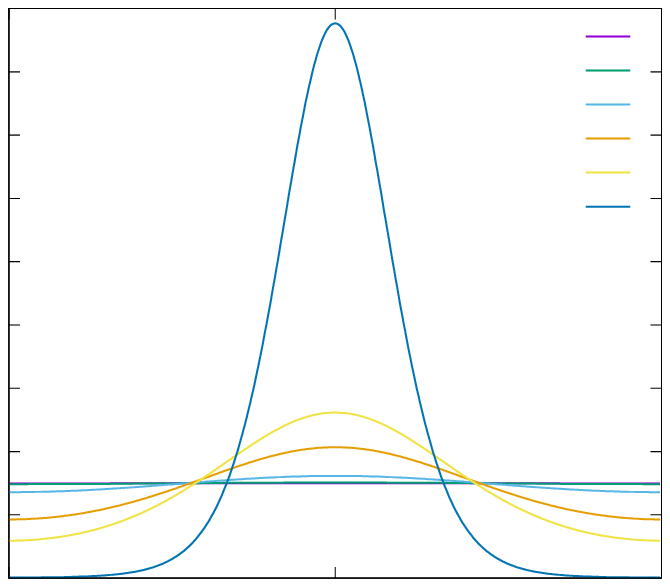}
\caption{We show the profile of the function $\phi(\chi)$ in the $\phi^4$-model (left)
as well as the energy density $\tilde{\varepsilon}$ of these solutions (right) for different values of $\alpha$}
\label{fig:phi4_solutions}
\end{center}
\end{figure} 

\begin{figure}[ht!]
\begin{center}
\input{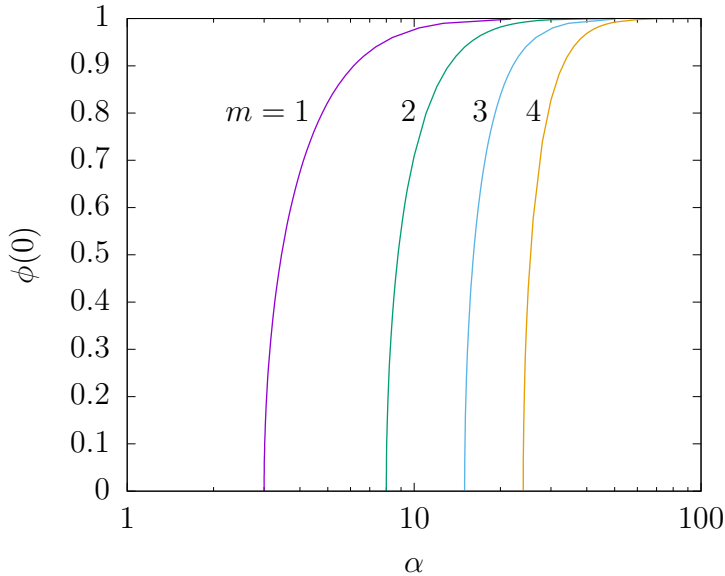} 
\caption{We show the value of $\phi(\chi=0)\equiv \phi(0)$ as function of $\alpha$ for the fundamental antikink solution ($m=1$), the kink-antikink ($m=2$) as well as the first excited antikink ($m=3$) and the first excited kink-antikink $(m=4)$, respectively. }
\label{fig:phi0_alpha}
\end{center}
\end{figure}

The antikinks of the $\phi^4$-model have been constructed in \cite{sudarsky} for $\alpha=4$. In Fig. \ref{fig:phi4_solutions}, we show the solutions to the model for different values of $\alpha$ as well as the corresponding energy densities $\tilde{\varepsilon}$. This figure demonstrates that
the value of $\phi(0)$ is a function of $\alpha$. In fact, when varying the value of $\alpha$, we observe that the solutions exist only to a minimal, non-vanishing value of $\alpha=\alpha_{min,1}$ which we find numerically to be $\alpha_{min,1}=3$. The reason for this is that the value of $\phi(\chi=0)$ is a decreasing function of $\alpha$ and for $\alpha$ sufficiently small, the kinks cease to exist.
This is also obvious when considering the plots of the energy density $\tilde{\varepsilon}$. For decreasing
$\alpha$, the energy density spreads over the $\chi$-direction and becomes equivalent to zero for $\alpha=\alpha_{min,1}$. This is related to the fact that $\alpha$ is the ratio between the width of the kink and the radius of the $S^3$ sphere. So, one would naturally expect that solutions exist only for radius larger than the width of the kink. This is shown in Fig. \ref{fig:phi0_alpha}, where we give the value of $\phi(\chi=0)$ in dependence of the coupling $\alpha$. Clearly for $\alpha < \alpha_{min,1}=3$ we find that $\phi(0)=0$ and hence $\phi(\chi)\equiv 0$. Increasing $\alpha$ from this value, we find that $\phi(0)$ increases up to $\phi(0)=1$, which
it reaches roughly when $\alpha={\cal O}(100)$. 

The limit of $\alpha \rightarrow \infty$ corresponds to the limit of flat, (3+1)-dimensional Minkowski space-time.
The existence of non-trivial, real scalar field solutions is forbidden by Derrick's theorem in this case.
The approach to this limit can also be seen in Fig. \ref{fig:phi4_solutions} and is confirmed by our numerics: for $\alpha\rightarrow\infty$, the value of $\phi(0)\rightarrow 1$, but at the same time, the function $\phi(\chi)$ has an increasing derivative at
$\chi=\pi/2$. In the limit $\alpha\rightarrow \infty$, we expect that $\phi(\chi)=1$ for $\chi\in[0:\pi/2[$, $\phi(\pi/2)=0$ and
$\phi(\chi)=-1$ for $\chi\in ]\pi/2:\pi]$ with an infinite derivative of $\phi$ and hence an infinite value of the energy density $\tilde{\varepsilon}$ at $\chi=\pi/2$.

\subsection{Kink-antikinks}
\label{subsection:KAK}
\begin{figure}[ht!]
\begin{center}
\input{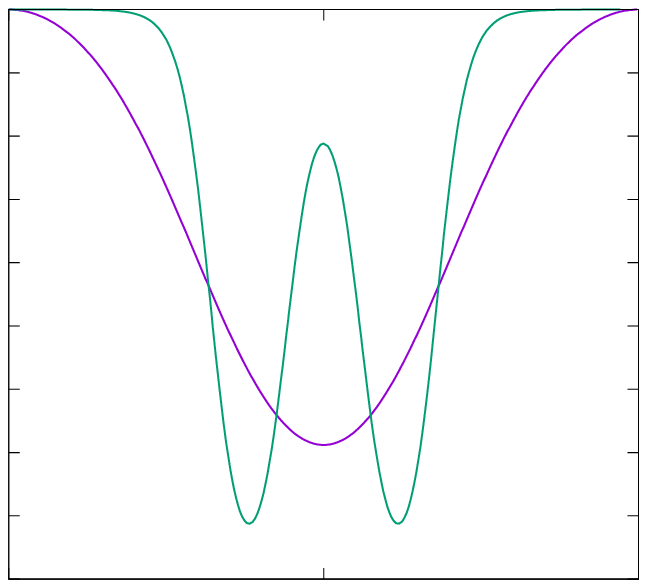}
\input{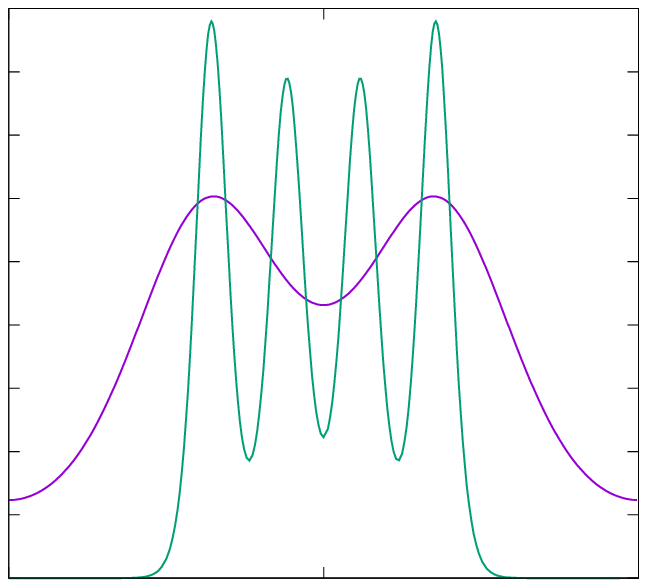}
\caption{We show the profile of a fundamental kink-antikink solution (left) as well as its energy density
$\tilde{\varepsilon}$ for $\alpha=10.0$. The profile function $\phi(\chi)$ of this solution has $m=2$ nodes (purple).
We also show the first excited kink-antikink solution and its energy density $\tilde{\varepsilon}$ for $\alpha=100.0$ (green). 
Note that this solution possesses $m=4$ nodes in the profile function $\phi(\chi)$.  }
\label{fig:KAK}
\end{center}
\end{figure}

Next to the fundamental kink solution described in (\ref{subsection:kinks}) for which $\phi(\chi=0)=-\phi(\chi=\pi)$, we have constructed solutions of (\ref{eq1}) fulfilling
$\phi(\chi=0)=\phi(\chi=\pi)$. These latter solutions have two nodes in the profile function $\phi(\chi)$. 
Using the notation of solutions of the $\phi^4$-model in (1+1) dimensions, we refer to these solutions as
{\it kink-antikinks}. In Fig. \ref{fig:KAK} we show an example of such a solution for $\alpha=10$ together with the energy density
$\tilde{\varepsilon}$.

The dependence of the value of the scalar field function $\phi(\chi=0)$ on $\alpha$ is shown in Fig. \ref{fig:phi0_alpha}.
Obviously, we need to choose the ratio between the width of the kink-antikink and the radius of the 3-sphere larger
to obtain these solutions. We find that at $\alpha_{\rm min,2}=8$, the value of $\phi(\chi=0)\rightarrow 0$ for these solutions, i.e. kink-antikinks exist only for $\alpha \geq 8$.

\subsection{$m$-excited kinks and kink-antikinks}
\label{subsection:mexcited}
\begin{figure}[ht!]
\begin{center}
\input{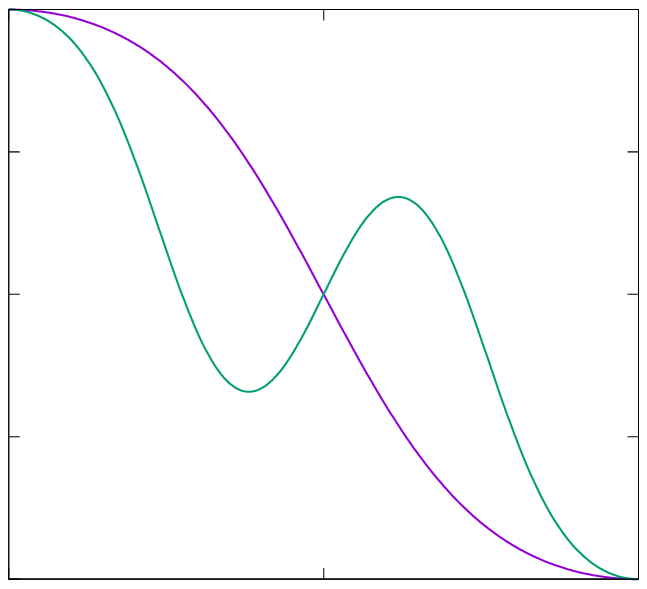}
\input{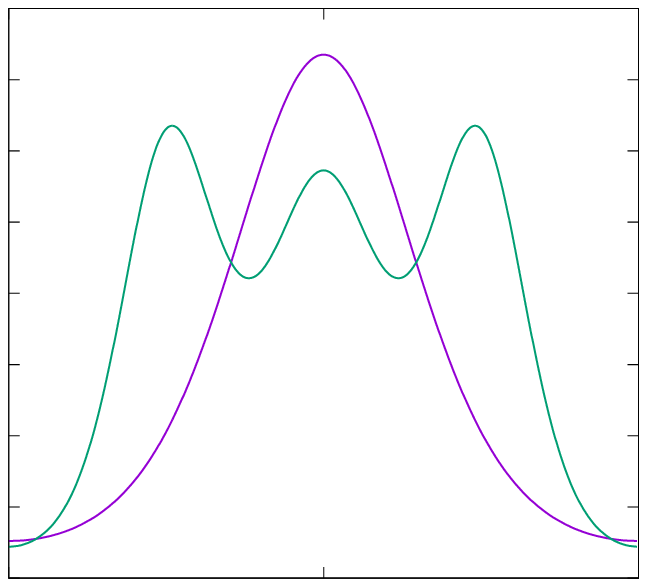}
\caption{We show the profile of a fundamental antikink solution (for $\alpha=5.0$, purple) together with a first excited antikink solution (for $\alpha=20.0$, green). We also give the energy density of these two solutions (right, same colour coding). }
\label{fig:excited}
\end{center}
\end{figure}

Next to the solutions desccribed in (\ref{subsection:kinks}) and (\ref{subsection:KAK}), we have constructed excited solutions of the former that differ in the number of nodes of the scalar field function $\phi(\chi)$. 
This is shown in Fig. \ref{fig:excited}, where we give examples of first excited antikink and the first excited kink-antikink
solutions, respectively.  Fig. \ref{fig:phi0_alpha} further demonstrates that first excited antikinks exist only for $\alpha \geq 15$, i.e. $\alpha_{\rm min,3}=15$, while first excited kink-antikinks exist only for $\alpha \geq 24$, i.e. $\alpha_{\rm min,4}=24$. The behaviour of $\alpha_{\rm min,m}$ can be explained easily when considering the 
small $\phi$ limit of (\ref{eq1}). In this limit, we can drop the $\phi^3$ term and the equation becomes linear.
Changing coordinates to $z=\cos\chi$, the equation then becomes a hypergeometric equation that has regular solutions
in the form of {\it Gegenbauer polynomials} if
\begin{equation}
\alpha_m=m(m+2) \ \ , \ \  m=1,2,3,4,.....  \ .
\end{equation}
This relation gives exactly the values of $\alpha_{\rm min,m}$ we find numerically when $\phi(\chi=0)\rightarrow 0$.

\section{Perturbations and their spectra}
\label{sec:stability}
The aim of our study of perturbations around the constructed solutions is two-fold: (a) 
since no-go theorems related to real scalar field solitons often require the additional property of stability, we will study whether
the solutions discussed above are linearly stable or unstable, and (b) the obtained spectra
are important in order to compare quantum fluctuations around the vacua of the model with those around solitons.

We use the following Ansatz for the perturbations about the solutions $\phi(\chi)$  to (\ref{eq1})~:
\begin{equation}
\label{perturbation}
\Phi(\vec{r},t)=\phi(\chi) + \Psi(\vec{r},t)   \ ,
\end{equation}
where $\Psi(\vec{r},t)$ is {\it a priori} complex-valued and we rescale $\Phi \rightarrow \eta \Phi$.
The equation for $\Psi$ then reads
\begin{equation}
\label{eq:perturbation}
\frac{1}{\sqrt{-g}}\partial_{\mu}\left(\sqrt{-g}\partial^{\mu}\Psi\right) - \alpha \left[ \phi^2 \left(2\Psi + \Psi^* \right) - \Psi + \phi \left(2 \vert\Psi\vert^2 + \Psi^2\right) + \vert\Psi\vert^2 \Psi \right] = 0 \ .
\end{equation}
In the following we will restrict our analysis to linear perturbations and hence neglect all 
terms of order ${\cal O}(\Psi^2)$ and higher. Note that these non-linear terms would mix the 
modes of the linear spectrum, a fact well known from the Fermi-Pasta-Ulam-Tsingou problem \cite{fput}.
We will use the following Ansatz for the linear perturbations 
\begin{equation}
\Psi(\vec{r},t)=F_1(\chi,\theta,\varphi) \exp(-i\omega t) + F_2(\chi,\theta,\varphi) \exp(i\omega t)
\end{equation}
where $F_i$, $i=1,2$ are assumed to be real-valued functions of the spatial coordinates and $\omega\in \mathbb{R}$ is chosen positive. Inserting this into the linearized version of (\ref{eq:perturbation}), we find
\begin{equation}
\frac{1}{\sqrt{-g}}\partial_{k}\left(\sqrt{-g}\partial^{k}F_i\right) - 
\alpha \left[\phi^2 \left(2F_i +  F_j \right)-F_i\right]=-\omega^2 F_i \ \ \ , \ \ \ i,j=1,2 \ , \ i\neq j \ , \ \ \ k=1,2,3  \ .
\end{equation}
This clearly demonstrated that negative and positive frequency modes couple, even for the
vacuum, i.e. absolute minium of the theory with $\phi^2\equiv 1$. 
In the following we will set $F_1= F_2\equiv F$ and use a product Ansatz for $F$ that reads~:
\begin{equation}
F(\chi,\theta,\varphi)=\psi(\chi) Y_{\ell \mu}(\theta,\varphi)  \ \ , \ \  
\end{equation}
where $Y_{\ell\mu}(\theta,\varphi)$, ${\ell}=0,1,2,3...$, $\mu=-{\ell},-{\ell}+1,....,{\ell}-1, {\ell}$ are the
spherical harmonics. Using the rescalings $t\rightarrow t/R_0$, $\omega\rightarrow R_0 \omega$ the equation for $\psi(\chi)$ then becomes~:
\begin{equation}
\label{eq:pert}
\psi''+ 2 \cot\chi \psi'- \alpha(3\phi^2 -1)\psi - \frac{\ell(\ell+1)}{\sin^2\chi} \psi = - \omega^2 \psi \ ,
\end{equation}
where the prime now and in the following denotes the derivative with respect to $\chi$. 
To solve this equation and find the eigenfrequencies $\omega$, we introduce a  new function
via $\tilde{\psi}=\sin^{-\ell}(\chi)\psi$ such that the equation reads
\begin{equation}
\label{eq:pert2}
\tilde{\psi}''+ 2 (\ell+1) \cot\chi \tilde{\psi}'- \alpha(3\phi^2 -1)\tilde{\psi} - \ell(\ell+2)\tilde{\psi} = - \omega^2 \tilde{\psi} \ .
\end{equation}
We now have to employ appropriate boundary conditions. These are
\begin{eqnarray}
\label{eq:bc2}
\tilde{\psi}'(\chi=0)=\tilde{\psi}'(\chi=\pi)=0  
\end{eqnarray}
and without loss of generality we will construct solutions that have $\tilde{\psi}(\chi=0)=1$.

In the following, we will investigate the perturbations about the vacua as well as the kink and anti-kink solutions. Related to the question of stability of the solutions, we will investigate whether $\omega^2 > 0$ ($\omega^2 < 0$) indicating that the solution is stable (unstable). 
We will also use the following notation: $\omega_{k,m,{\ell}}$ will denote the energy eigenvalues of 
the $k$-th excited perturbation with angular number $\ell$, i.e.
a solution to (\ref{eq:pert}) with $k$ nodes, of the scalar field solution of (\ref{eq:em_scalar}) with $m$ nodes associated to the ${\ell}$-th spherical harmonic. In this notation, $m=0$ corresponds to the vacua, $m=1$ the (anti)kink, $m=2$ the kink-antikink and higher $m$ to the excitations, respectively.

\subsection{Perturbations about the vacua ($m=0$)}
For the vacua $\psi(\chi)\equiv \pm 1$, i.e. the global minima of the potential, we obtain the equation
\begin{equation}
\label{eq_per1}
\psi''+ 2\cot\chi \psi'- 2\alpha \psi -\frac{\ell(\ell+1)}{\sin^2 \chi} \psi= -\omega^2 \psi   \ .
\end{equation}
This equation is similar to that discussed in the case of a conformally coupled scalar field
in an Einstein universe \cite{ford2} and the general solutions to the equation are
\begin{equation}
\psi_{k,\ell}(\chi)=\sin^{\ell}(\chi)C_{k-\ell}^{\ell+1}(\cos\chi)  \ \ , \ \  k=0,1,2,...  \ ,
\end{equation}
where the $C_{k-\ell}^{\ell+1}$ are the Gegenbauer functions and $\ell$ is now restricted by $\ell\leq k$. 
The corresponding {\it discrete} eigenfrequencies read~: 
\begin{equation}
\label{spectrum_gegenbauer}
\omega^2_{k,0,{\ell}}=k(k+2) + 2\alpha  \ \ \ , \ \ \  k=0,1,2,3,...  \ ,
\end{equation}
which do not depend on ${\ell}$. Hence, all multipoles have the same eigenfrequency such
that for a fixed value of $k$ the number of degenerate modes is $(k+1)^2$. 

Note that $\omega^2_{k,0,{\ell}}$ is always positive and always larger than $2\alpha$. This indicates that the trivial solutions $\phi(\chi)=\pm 1$, i.e. the vacua, are stable.

\subsection{Perturbations about the (anti)kink ($m=1$)}
\label{sub:pert_kink}

For the (anti)kink solutions discussed in \ref{subsection:kinks}, the equation (\ref{eq:pert}) has to be solved numerically.  We find a discrete spectrum of modes that depends on $\alpha$. In the following, we will first discuss our results for the case ${\ell=0}$, i.e. the monopole contribution to the perturbations, and then comment on $\ell >  0$, i.e. higher multipoles. 

\subsubsection{The monopole ($\ell=0$)} 
The lowest mode with $k=1$ has a profile for $\psi$ that resembles the (anti)kink solution itself, while higher modes, $k > 1$, resemble the radially excited solutions discussed in \ref{subsection:mexcited}. Note that a mode with $\psi\equiv {\rm constant}\neq 0$ does no exist in
this case as is apparent from (\ref{eq:pert}), i.e. the first mode fulfilling the periodic boundary conditions is a solution with one node, $k=1$.

\begin{figure}[ht!]
\begin{center}
\input{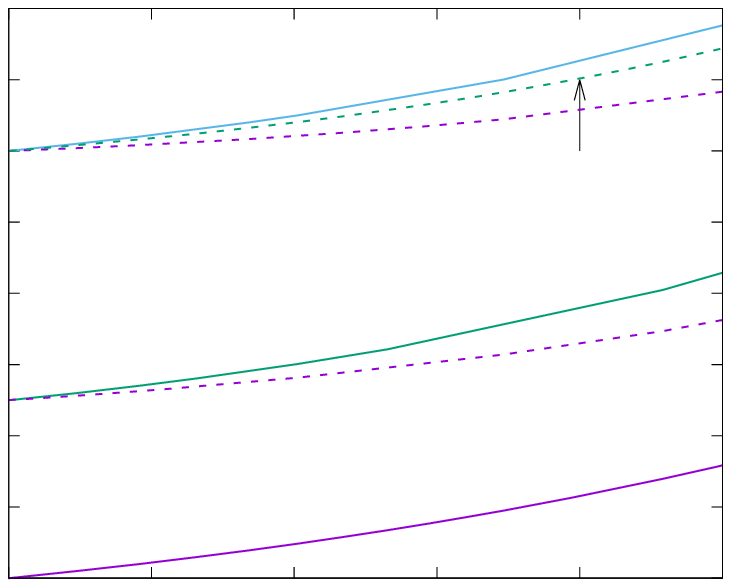}
\input{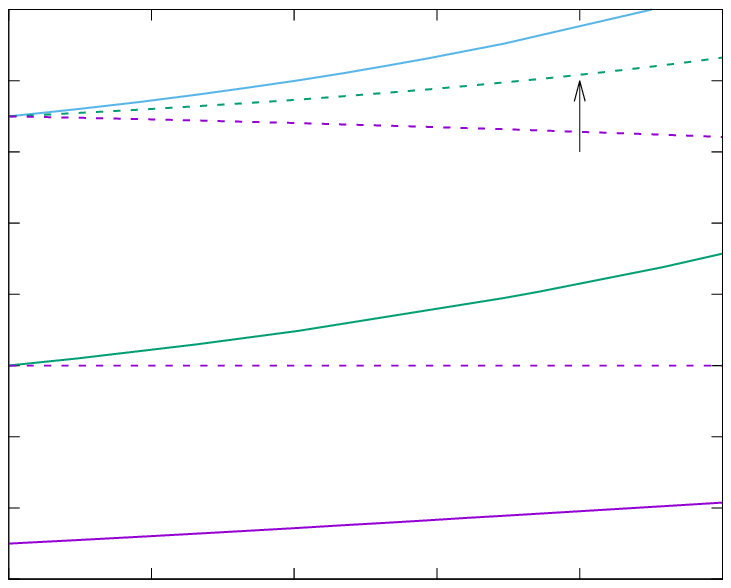}
\caption{We show the lowest eigenfrequencies $\omega^2_{k,m,\ell}$ of the $k$-th perturbation 
about the $m$-th soliton with angular quantum number $\ell$ in dependence on $\alpha$. {\it Left:} The value of $\omega_{k,1,0}^2$ for the monopole ($\ell=0$) perturbations (solid) about the (anti)kink solution ($m=1$)  for
$k=1,2,3$. We also give the value of $\omega^2_{k,1,1}$ of the dipole ($\ell=1$) perturbation for $k=1,2$ and $\omega^2_{1,1,2}$ of the quadrupole ($\ell=2$) perturbation for $k=1$ (dashed).
{\it Right:} The values of  $\omega_{k,2,0}^2$ for the monopole ($\ell=0$) perturbation about the kink-antikink solution ($m=2$) for $k=1,2,3$ (solid).  We also give the value of $\omega^2_{k,2,1}$ of the dipole ($\ell=1$) perturbation for $k=1,2$ and $\omega^2_{1,2,2}$ of the quadrupole ($\ell=2$) perturbation for $k=1$ (dashed). }
\label{fig:spectrum_1_k}
\end{center}
\end{figure}

For $\alpha\rightarrow \alpha_{min,1}=3$, we find $\omega_{1,1,0}^2\rightarrow 0$, $\omega_{2,1,0}^2\rightarrow 5$, $\omega_{3,1,0}^2\rightarrow 12$, which indicates that 
\begin{equation}
\omega_{k,1,0}^2 \rightarrow k(k+2)-3  \ \ \ {\rm for} \ \  \alpha\rightarrow 3  \ . 
\end{equation}
Moreover, Fig. \ref{fig:spectrum_1_k} (left) demonstrates that the value of $\omega_{k,1,0}^2$ increases when increasing $\alpha$ from its limiting value $\alpha_{min,1}=3$.  
Hence, the eigenvalues $\omega_{k,1,0}^2$ of all $\ell=0$ perturbations about the (anti)kink solution are positive. 


\subsubsection{The dipole ($\ell=1$) and quadrupole ($\ell=2$)} 
We have studied the dipole ($\ell=1$) and quadrupole ($\ell=2$)  contribution for different values
of $k$. Our results for the lowest modes are shown in Fig.\ref{fig:spectrum_1_k}. 
As is obvious from this figure, {\it the degeneracy of the eigenfrequencies with respect to the angular number $\ell$ no longer exists in the presence of the (anti)kink}. We find that $\omega^2_{k,1,\ell}$ for fixed $k$ increases with increasing $\ell$, see the values for $\omega^2_{1,1,\ell}$ (dashed lines). Moreover,
the eigenfrequencies $\omega^2_{1,1,2}$ are comparable in value to the eigenfrequencies
$\omega^2_{2,1,1}$ and $\omega^2_{3,1,0}$ and in the limit $\alpha\rightarrow \alpha_{\rm min,1}=3$ tend to the same value.  {\it This suggests that for a fixed $m$, the eigenfrequencies $\omega^2_{k,m,\ell}$
with $k+\ell$ constant are comparable in value and degenerate for $\alpha\rightarrow\alpha_{min,m}$}.  
 Similar to the monopole discussed above, the eigenfrequencies for all $k$ and $\ell$ that we have studied are positive and since their values increase with both $k$ and $\ell$, we find that {\it the (anti)kink
is linearly stable under general perturbations}. 

\subsection{Perturbations about the kink-antikink ($m=2$)}
\label{sub:pert_kink_antikink}
Again, we will first discuss the monopole perturbation, $\ell=0$, and then comment on $\ell > 0$.

\subsubsection{The monopole ($\ell = 0$)} 

In Fig. \ref{fig:spectrum_1_k} (right), we give the dependence of the eigenvalues $\omega_{k,2,0}^2$, $k=1,2,3$ on $\alpha$.
In contrast to the perturbations about the vacuum and the (anti)kink, respectively, we find one mode with negative eigenvalue. 
This is the lowest mode, $k=1$, for which the profile $\psi_1$ possesses one node and eigenvalue $\omega_{1,2,0}^2 < 0$.
For $\alpha\rightarrow \alpha_{min,2}=8$, we find that $\omega_{1,2,0}^2 \rightarrow -5$. 
This eigenvalue increases when increasing $\alpha$, but stays negative for all values of $\alpha$ that we have investigated. We find e.g. that $\omega_{1,2,0}^2\approx -2.12$ for $\alpha=100$ and our numerical results further indicate that $\omega_{1,2,0}^2 \rightarrow 0$ for $\alpha\rightarrow \infty$.  
On the other hand, the eigenvalues of higher modes -- in Fig. \ref{fig:spectrum_1_k} (right) shown for $k=2,3$ -- are positive for all values of $\alpha \geq \alpha_{min,2}=8$ with $\omega_{2,2,0}^2 \rightarrow 0$ for $\alpha\rightarrow \alpha_{min,2}$. 
We conclude that the kink-antikink solution has (at least) one unstable mode and hence is {\it linearly unstable}.  

\subsubsection{The dipole ($\ell=1$) and quadrupole ($\ell=2$)} 
Considering the dipole and quadrupole contributions of the perturbations, we find again that
the degeneracy with respect to $\ell$ does not longer exist in the presence of the kink-antikink.
Moreover, a few other interesting things appear. The first thing to note is that the dipole
contribution of the $k=1$ perturbation about the kink-antikink has eigenfrequency $\omega^2_{1,2,1}\equiv 0$ for all values of $\alpha$ that we have studied. Hence, this perturbations corresponds to
a {\it zero mode} of the system. Moreover, similar to the (anti)kink case, the eigenfrequencies
$\omega^2_{3,2,0}$, $\omega^2_{2,2,1}$ and $\omega^2_{1,2,2}$
are comparable in value and become degenerate in value for $\alpha\rightarrow \alpha_{min,2}=8$.
This strengthens the claim made above for $m=1$ that for a fixed $m$, the eigenfrequencies $\omega^2_{k,m,\ell}$
with $k+\ell$ constant are comparable in value and degenerate for $\alpha\rightarrow\alpha_{min,m}$. 
There is one further interesting point that is new in comparison to perturbations about the (anti)kink.
We find that the eigenfrequency $\omega^2_{1,2,2}$, i.e. the quadrupole contribution to the $k=1$ perturbations has {\it decreasing values for $\alpha$ increasing} - in contrast to all the other perturbations that we have discussed up to here.

\subsection{Perturbations for $m > 2$ and/or $\ell > 2$ and/or $k > 3$}
In order to strengthen the claims about the eigenfrequency spectra, we have studied several other values of $k$, $m$, $\ell$. The values of $\omega^2_{k,m,\ell}$ for $\alpha=\alpha_{min,m}+1$  and
$\alpha=\alpha_{min,m}+2$, respectively, and  
all possible combinations for $k\in [1:3]$, $m\in [1:3]$ and $\ell \in [0:3]$ are given in 
Table \ref{table:omega_m_k_l}. For the perturbations about the (anti)kink ($m=1$), we find that
higher $\ell$ follow the pattern identified before. For all values of $k,\ell$, we find that
$\omega^2_{k,1,\ell}$ is positive and increases with $\alpha$. 

When considering perturbations about the kink-antikink ($m=2$), we find that the eigenfrequencies
for $k+\ell$ constant have the lowest value for the highest possible $\ell$. This indicates that
multipole perturbations along the $\theta$, $\varphi$ directions on the $S^3$ cost less energy than the perturbations in the $\chi$-direction.  We also find that for $\ell > k$ the eigenfrequency 
decreases with increasing $\alpha$. 

A very similar pattern as that described above appears for the case $m=3$, however, there are some
small differences. First, we find a number of negative eigenfrequencies: $\omega^2_{k,3,\ell} \leq 0$
for $k < m$ and $\ell < m$ and the equality holding for $k=m-1=2$, $\ell=1$. Moreover, the statement
that for $k+\ell$ constant the highest $\ell$ has the lowest eigenfrequency is only true for
$k+\ell \leq 3$. For $k+\ell=4$, we find that the $k=\ell=2$ perturbation has lower value of the eigenfrequency than the $k=1,\ell=3$ perturbation. Moreover, all perturbations with $\ell \geq k$ 
have now decreasing value of the eigenfrequency when increasing $\alpha$.

\begin{figure}[ht!]
\begin{center}
\input{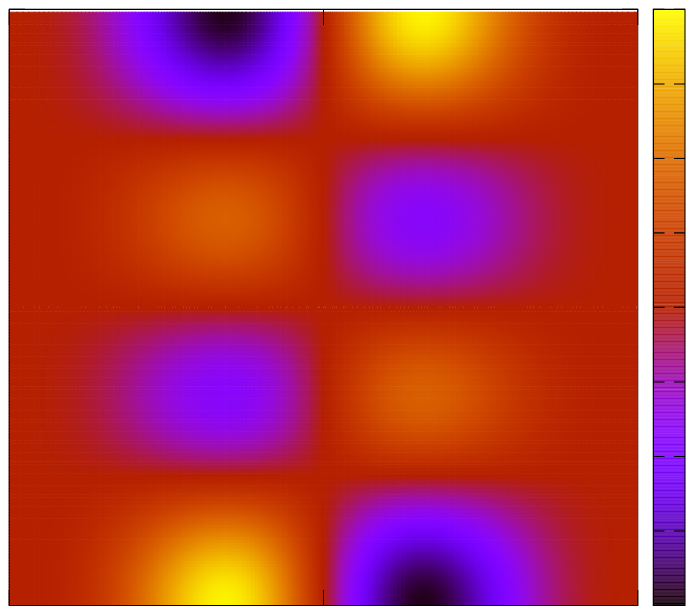}
\input{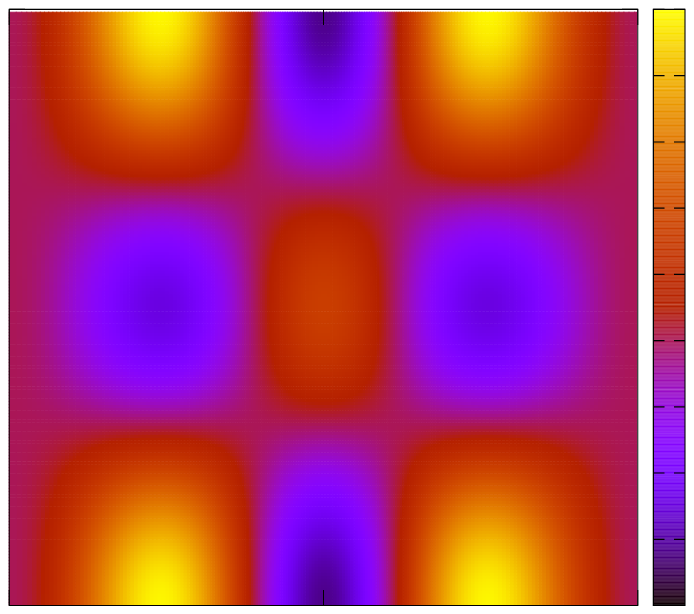}
\caption{We show the value of $\vert\psi\vert$ of the perturbations about the kink solution ($m=1$) for $\alpha=4$.
{\it Left:} The octupol contribution ($\ell=3$) to the first perturbation ($k=1$) which
has $\omega^2_{1,1,3}\approx 21.61$.
{\it Right:} The quadrupole contribution ($\ell=2$) to the second perturbation ($k=2$) which
has $\omega^2_{2,1,2}\approx 22.33$.
 }
\label{fig:absolute_psi_higher_ell}
\end{center}
\end{figure}

In Fig.\ref{fig:absolute_psi_higher_ell}, we show $\vert\psi\vert$ for two perturbations about the
(anti)kink solution for $k+\ell$ constant, in this case $k+\ell=4$. We have chosen the
octupol contribution ($\ell=3$) of the first excitation ($k=1$) as well as the quadrupole
contribution ($\ell=2$) of the second excitation ($k=2$). These perturbations are obviously
very different in nature, but have comparable eigenfrequencies at $\alpha=4$.

\begin{table}
\centering
\begin{tabular}{|c|c|c|l|l|}
\hline
$m$ & $k$ & ${\ell}$ & $\omega^2_{k,m,\ell}$ for $\alpha_{ min,m}+1$ & $\omega^2_{k,m,\ell}$ for $\alpha_{min,m}+2$\\
\hline\hline
$1$ & $1$ & $0$ & $1.90$ & $3.65$ \\
\hline
$1$ & $1$ & $1$ & $6.28$ & $7.61$ \\
\hline
$1$ & $2$ & $0$ & $ 7.05$ & $9.14$ \\
\hline
$1$ & $1$ & $2$ & $12.89$ & $13.91$ \\
\hline
$1$ & $2$ & $1$ & $13.65$ & $15.38$ \\
\hline
$1$ & $3$ & $0$ & $14.01$ & $16.02$ \\
\hline
$1$ & $1$ & $3$ & $21.61$ & $22.40$ \\
\hline
$1$ & $2$ & $2$ & $22.33$ & $23.79$ \\
\hline
$1$ & $3$ & $1$ & $22.77$ & $24.58$ \\
\hline
$1$ & $2$ & $3$ & $33.08$ & $34.32$ \\
\hline
$1$ & $3$ & $2$ & $33.54$ & $35.16$ \\
\hline
$1$ & $3$ & $3$ & $46.33$ & $47.79$ \\
\hline
\hline
$2$ & $1$ & $0$ & $-4.22$ & $-3.74$ \\
\hline
$2$ & $1$ & $1$ & $0$ & $0$ \\
\hline
$2$ & $2$ & $0$ & $1.89$ & $3.59$ \\
\hline
$2$ & $1$ & $2$ & $6.64$ & $6.34$ \\
\hline
$2$ & $2$ & $1$ & $7.95$ & $8.93$ \\
\hline
$2$ & $3$ & $0$ & $9.04$ & $11.10$ \\
\hline
$2$ & $1$ & $3$  & $15.46$ & $14.99$ \\
\hline
$2$ & $2$ & $2$ & $16.36$ & $16.86$ \\
\hline
$2$ & $3$ & $1$ &  $17.36$ & $18.78$ \\
\hline
$2$ & $2$ & $3$  & $ 26.99$ & $27.14$ \\
\hline
$2$ & $3$ & $2$ &  $27.83$ & $28.79$ \\
\hline
$2$ & $3$ & $3$  &  $40.43$ & $41.03$ \\
\hline\hline
$3$ & $1$ & $0$  & $-11.66$ & $-11.52$ \\
\hline
$3$ & $1$ & $1$  & $-7.24$ & $-7.47$ \\
\hline
$3$ & $2$ & $0$ & $-5.95$ & $-5.23$ \\
\hline
$3$ & $1$ & $2$  & $-0.39$ & $-0.75$ \\
\hline
$3$ & $2$ & $1$ & $0$ & $0$ \\
\hline
$3$ & $3$ & $0$ & $1.91$ & $3.62$ \\
\hline
$3$ & $2$ & $2$ & $8.56$ & $8.18$ \\
\hline
$3$ & $1$ & $3$  & $8.60$ & $8.20$ \\
\hline
$3$ & $3$ & $1$ & $9.76$ & $10.54$ \\
\hline
$3$ & $2$ & $3$  & $19.41$ & $18.85$ \\
\hline
$3$ & $3$ & $2$ & $20.07$ & $20.25$ \\
\hline
$3$ & $3$ & $3$  & $32.68$ & $32.48$ \\
\hline
\end{tabular}
\caption{The value of $\omega^2_{k,m,\ell}$ for different choices of $m$, $k$, $\ell$  at
$\alpha_{min,m}+1$ and $\alpha_{min,m}+2$, respectively. For $m=1$, $m=2$, $m=3$ the value is 
$\alpha_{min,1}=3.0$,
$\alpha_{min,2}=8.0$, $\alpha_{min,3}=15.0$, respectively.}
\label{table:omega_m_k_l}
\end{table}

\section{Conclusions}
\label{sec:conclusions}
We have demonstrated that the no-go theorem for real, static, stable scalar field solitons in (3+1)-dimensions in a curved space-time background do not extend to a space-time that is spatially compact. We have constructed (anti)kink and kink-antikink solutions
in an Einstein universe background and have shown that the (anti)kink solutions are linearly stable. Moreover, $m$-excited solutions
do exist that possess $m$ nodes in the scalar field profile. 

The study of the linear perturbations about the solutions has shown a number of interesting features.
The degeneracy with respect to the angular number $\ell$ that is present for the perturbations
about the vacua does no longer exist in the presence of the solitonic-like objects.
Moreover, we find that for an $m$-kink the eigenfrequencies associated to the perturbations
$\omega^2_{k,m,\ell}\leq 0$, hence indicating instabilities, whenever $k < m$
and -- at the same time -- $\ell < m$. Since the lowest order perturbation about the (anti)kink ($m=1$) is $k=1$, there are
no negative eigenfrequency perturbations and {\it the (anti)kink is linearly stable}.
For the particular case that $k=m-1$ and $\ell=1$ we find that the corresponding {\it eigenfrequency
$\omega^2_{m-1,m,1}$ has value zero}. Hence, the corresponding perturbations correspond to {\it zero modes} of the model.  

The eigenfrequency spectra and the corresponding perturbations $\psi$ are important in
the computation of quantum effects in given space-times, especially in those with a non-vanishing
curvature, see e.g \cite{ford2,ford1}. Often compact space-times are considered rendering the
eigenfrequency spectra discrete. Here, we have studied the example of a positively curved, spatially compact space-time given by the Einstein universe. While this static universe is not the universe
we observe, it has features similar to the latter: it contains a perfect fluid with positive energy density
and vanishing pressure (ordinary matter) as well as a positive cosmological constant. In  computations
of quantum effects it is often considerd a good approximation to the observed universe in sufficiently
small time intervals. What we have demonstrated in this paper is that the presence of solitonic-like
objects (in the form of (anti)kinks) modifies the eigenfrequency spectra of a possible
quantum perturbation (given by the field $\psi$) considerably. Since the universe
is supposed to have gone through a number of phase transitions in the early universe 
during which topological defects, i.e. localized solitonic-like objects, (could) have formed, computations
of quantum effects (e.g. the Casimir effect) should certainly take these into account.

Let us finally remark that in this paper we have only studied static solutions and demonstrated (see Appendix) that a full back-reacted solution with
space-time symmetries equivalent to that of the Einstein universe cannot be static. However, one can consider the 
non-linear dynamics of the scalar field itself on a static space-time background. The first possibility would then be to 
study the dynamical decay of the unstable solutions in our model and get an estimate for the scalar radiation emitted in this process.
Moreover, the simulation of the scattering of the
solutions in our model will be interesting in order to see how the fact that the background is not (asymptotically) Minkowski will 
influence the scattering properties of the diverse solutions.  Since the model at hand is not integrable, we would expect internal modes
to be excited in the scattering. Moreover, the study of the non-linear dynamics of the full back-reacted system would be interesting in order to understand how the stability
of the solutions is influenced by a dynamical space-time and how and whether -- next to scalar radiation -- gravitational radiation would be 
emitted in the decay or scattering of the solitons. This is currently under investigation.

\vspace{1cm}

{\bf Acknowledgements} 
 BH would like to thank FAPESP for financial support under
grant number {\it 2019/01511-5}. GL would like to thank FAPES for the financial support under the EDITAL CNPq/FAPES N$^o$ 22/2018.
CFSP would like to thank FAPES for financial support under grant N$^o$ 98/2017. We would like to thank FAPES for support
under grant N$^o$ 0447/2015. 

\clearpage


\clearpage

\section{Appendix: No static universe for scalar matter}
Let us assume the metric to be of the FLRW type
\begin{equation}
ds^2 = -dt^2 + R(t)^2 \left[d\chi^2 +\sin^2\chi\left(d\theta^2 + \sin^2\theta d\varphi^2\right)\right] \ 
\end{equation}
such that the non-vanishing components of the Einstein tensor read~:
\begin{equation}
G_{tt} = 3 \left(\frac{\dot{R}^2}{R^2} + \frac{1}{R^2}\right) \ \ \ , \ \ \ 
G_{\chi\chi} = - \left( 2\ddot{R} R + \dot{R}^2 + 1\right)  \ \ , \ \ 
G_{\theta\theta}=\sin^2 \chi G_{\chi\chi} \ \ \ , \ \ \   G_{\varphi\varphi}=\sin^2\chi \sin^2 \theta \ G_{\chi\chi} \ ,
\end{equation}
and the dot denotes the derivative with respect to $t$.  Using the components of the energy-momentum tensor of a static scalar
field given in (\ref{eq:em_scalar}) and replacing $R_0 \rightarrow R(t)$, the relevant components of the Einstein equation $G_{\mu\nu} + \Lambda g_{\mu\nu}=8\pi G T_{\mu\nu}$ are
\begin{eqnarray}
  3\left(\frac{\dot{R}^2}{R^2} + \frac{1}{R^2}\right) - \Lambda  &=& 8\pi G\left[\frac{1}{2} \frac{\phi'^2}{R^2} + V(\phi) \right]  \\
  - \left( 2\frac{\ddot{R}}{R} + \frac{\dot{R}^2}{R^2} + \frac{1}{R^2}\right) + \Lambda &=& 8\pi G \left[\frac{1}{2}\frac{\phi'^2}{R^2} - V(\phi)\right]  
\end{eqnarray}
Combining these two equations for the case of a static universe, $\dot{R}=\ddot{R}\equiv 0$, we find that 
$4\pi G \phi'^2=1$, which implies $\phi(\chi)\sim \chi + {\rm constant}$. This solution is not compatible with the
periodicity condition and hence, no static universe exists when choosing a static scalar field as the given matter source.

 \end{document}